\documentclass[Afour,sagev,times]{sagej}

\usepackage{moreverb,url}

\usepackage[colorlinks,bookmarksopen,bookmarksnumbered,citecolor=red,urlcolor=red]{hyperref}

\usepackage{amsmath}
\usepackage[mathscr]{euscript}
\usepackage{multirow}
\usepackage{epstopdf}% To incorporate .eps illustrations using PDFLaTeX, etc.
\usepackage[caption=false]{subfig}% Support for small, `sub' figures and tables
\usepackage{caption}

\usepackage{graphicx}
\usepackage{adjustbox}
\usepackage{natbib}
\makeatletter% @ becomes a letter
\def\NAT@def@citea{\def\@citea{\NAT@separator}}% Suppress spaces between citations using natbib.sty
\makeatother% @ becomes a symbol again

\theoremstyle{plain}% Theorem-like structures provided by amsthm.sty

\theoremstyle{definition}

\theoremstyle{remark}

\newcommand\BibTeX{{\rmfamily B\kern-.05em \textsc{i\kern-.025em b}\kern-.08em
		T\kern-.1667em\lower.7ex\hbox{E}\kern-.125emX}}

\def\volumeyear{2020}

\begin{document}

\runninghead{Lee and Boone}

\title{Applications of Bayesian Dynamic Linear Models to Random Allocation Clinical Trials}
%\itshape{SAGE Publications}}

\author{Albert H. Lee III\affilnum{1}, Edward L. Boone\affilnum{1}, Roy T Sabo\affilnum{2}, Erin Donahue\affilnum{3}}

\affiliation{\affilnum{1}Department of Statistical Sciences and Operations Research, Virginia Commonwealth University, Richmond, Virginia 23284 USA;\\
\affilnum{2}Department of Biostatistics, Virginia Commonwealth University, Richmond, Virginia, 23284 USA;\\
\affilnum{3}Levine Cancer Institute, Charlotte, North Carolina, 28204 USA;}

\corrauth{Albert H. Lee III,
	Department of Statistical Sciences and Operations Research, Virginia Commonwealth University, Richmond, Virginia 23284 USA;}

\email{leeah2@vcu.edu}

\begin{abstract}
	Random allocation models used in clinical trials aid researchers in determining which of a particular treatment provides the best results by reducing bias between groups. Often however, this determination leaves researchers battling ethical issues of providing patients with unfavorable treatments. Many methods such as Play the Winner and Randomized Play the Winner Rule have historically been utilized to determine patient allocation, however, these methods are prone to the increased assignment of unfavorable treatments. Recently a new Bayesian Method using Decreasingly Informative Priors has been proposed by \citep{sabo2014adaptive}, and later \citep{donahue2020allocation}. Yet this method can be time consuming if MCMC methods are required. We propose the use of a new method which uses Dynamic Linear Model (DLM) \citep{harrison1999bayesian} to 
	to increase allocation speed while also decreasing patient allocation samples necessary to identify the more favorable treatment. Furthermore, a sensitivity analysis is conducted on multiple parameters. Finally, a Bayes Factor is calculated to determine the proportion of unused patient budget remaining at a specified cut off and this will be used to determine decisive evidence in favor of the better treatment.
	
%This paper describes the use of the \LaTeXe\
%\textsf{\journalclass} class file for setting papers to be
%submitted to a \textit{SAGE Publications} journal.
%The template can be downloaded \href{http://www.uk.sagepub.com/repository/binaries/SAGE LaTeX template.zip}{here}.
\end{abstract}

\keywords{Bayes Factor, Dynamic Linear Model, Random Allocation, Clinical Trials, Time Series}

\maketitle

\section*{Introduction}
Clinical trials are controlled methods by which researchers may \lq\lq{obtain sound scientific evidence for supporting the adoption of new therapies in clinical medicine\rq\rq} \cite{zelen1969play}.  Clinical trials are defined by \cite{zelen1969play} \lq\lq{to consist of at least two groups of patients who are as similar as possible except for the administered treatment whereby the groups are decided through randomization\rq\rq}. Extensive research has be done in the randomization of clinical trials. The most common approach consists of equally allocating the same number of subjects to two treatments. Yet, \cite{ivanova2003play} pointed out this method suffers from ethical issues provided one drug is superior, while also possessing a less than adequate parameter estimating ability. Thus one would like to be able to sequentially allocate participants in such a way that the randomization remains preserved, while also skewing participants to the better treatment. This is known as adaptive allocation. 

Methods for adaptively allocating subjects between treatments include the earliest works conducted by \cite{thompson1933likelihood} who was the first to look at what has become known as the adaptive design. Additional adaptive design works include those of \cite{anscombe1963sequential}, as well as \cite{colton1963model}. Likewise, \cite{robbins1952some} contribution to adaptive allocation led to the Play the Winner Rule which allocates patients to future successful trials based on the success of one trial or failure of the other (see \cite{zelen1969play} for details). While \cite{rosenberger1999randomized} suggests this method can be a useful substitute for equal allocation, he indicates there is lower power when compared to equal allocation models. The Play the Winner Rule was modified by \cite{wei1978randomized} into the Randomized Play the Winner Rule. Further works include those of \cite{ivanova2003play} as well as \cite{wei1979generalized}. \cite{sabo2017optimal} examined a comparison between the optimal design of \cite{rosenberger2001optimal} and the approach of \cite{thall2007practical} for binary outcomes using a Bayesian approach. Likewise, \cite{sabo2014adaptive} used a Bayesian approach to create what he termed  \lq\lq{Decreasingly Informative Prior\rq\rq} information to examine how adaptive allocation performed on binary variables. Each of these aforementioned methods are a type of urn randomization method and as such, each of these methods have binary responses leading to proportional allocation. For more on urn randomization methods and their properties see \cite{wei1988properties}. 

Another method used is the Bayesian Adaptive Design in which assignment of either treatment or control is conducted through adaptive allocation. Extensive work has been done in this area including the works of \cite{connor2013bayesian} who determined the method provided \lq\lq{improved patient outcomes and increased power\rq\rq} along with a \lq\lq{lower expected sample size\rq\rq} in a three arm trial in which one treatment was actually better than the others. Another area this method has been used is in the Recurrent Glioblastoma trial conducted by \cite{trippa2012bayesian}. They concluded \lq\lq{the use of Bayesian adaptive designs in glioblastoma trials would result in trials requiring substantially fewer overall patients, with more patients being randomly assigned to efficacious arms\rq\rq}. The lung cancer study of \cite{zhou2008bayesian} utilized a probit model and was found, along with a suitable early stopping rule, to be an ethical design which can be used to improve personalized medicine. Bayesian adaptive design has been used to design a trial to analyze acute heart failure syndromes by \cite{collins2012bayesian}. They determined this type of \lq\lq{clinical trial represents an innovative and potentially paradigm-shifting method of studying personalized treatment options for AHFS\rq\rq}.

Regardless of the previously mentioned designs chosen, the $y$ patients enter a random allocation study sequentially at different times. Thus patients entering a random allocation study may be considered a set of time series measurements. Furthermore, throughout the trial there will be a total of $\mathscr{N}$ patients. Let $\mathscr{T}$ 
be the index set for patient $y_{t}$ measured in a total of $\mathscr{N}$ patients. Because these $y_{t}$ patients enter the allocation study sequentially, more information regarding allocation to the better treatment is known at, say, patient $y_{10}$, than at patient $y_{9}$.  This allows the researchers to learn more information about treatment effectiveness as patients enter the study. However, with the Bayesian allocation designs, this information becomes a Bayesian Learning Design; as the information is updated, the Bayesian design learns which treatment is better.

Bayesian adaptive designs use Bayesian updating methods to allocate subjects to treatments. The ability of using the posterior as the prior through repeated updating makes these Bayesian methods \lq\lq{a natural framework for making decisions based on accumulating data during a clinical trial\rq\rq} \cite{thall2007practical}. Furthermore, this updating ability provides as a fortuitous side effect, according to \cite{berry2006bayesian} \lq\lq{the ability to quantify what is going to happen in a trial from any point on (including from the beginning), given the currently available data\rq\rq}.

\section*{Bayesian Methods}
%The first discussion of Bayesian methodologies is a 1763 analysis by Reverend Thomas Bayes (see \cite{mr1763essay} for more details). Numerous works utilizing Bayesian methodologies have been written including the general theoretical overview using Bayesian sampling methods provided by \cite{gelman1995bayesian}. The works of \cite{lee2012bayesian} combines the use of the $R$ programming language with the theoretical overviews given by the aforementioned authors. A general introduction to Bayesian methods may be found in \cite{petris2009dynamic} chapter 1. The works of \cite{berger2013statistical} provide a general discussion of Bayesian Loss functions. 

The basic premise surrounding Bayesian methods is known as Bayes rule, named after Rev. Thomas Bayes who postulated the probability of some unknown parameter $\theta$, given the corresponding observations $y$, was simply the ratio of the probability of the joint density function $p(\theta,y)$ to the probability we observe the value $y$. Mathematically speaking this is
\begin{center}
	\begin{equation} %\label{eq:3}
	p(\theta | y) = p(\theta)p(y| \theta)/p(y)
	\end{equation} \label{eq:1}
\end{center}
where $p(\theta | y)$ is now the posterior, or updated distribution for $\theta$ given some $y$ and
\begin{center}
	\begin{equation} %\label{eq:4}
	p(\theta)p(y| \theta) = p(\theta,y)
	\end{equation} \label{eq:2}
\end{center}

According to \cite{gelman1995bayesian} $p(\theta)$ is some prior distribution of parameters and $p(y| \theta)$ is the sampling distribution such that conditioning on the known $y$ data will lead to the posterior distribution (See \cite{gelman1995bayesian} for more details.) This idea has been extended upon by \cite{petris2009dynamic} for time series data. 
The learning ability available through this updating process in these Bayesian methods has been extended by \cite{harrison1999bayesian} using Dynamic Linear Models (DLM). The DLM uses this Bayesian Learning Process to update and forecast the $y$ observations such that
\begin{eqnarray} %\label{eq:5} \label{eq:5}
\boldsymbol{Y_{t}} &=& \boldsymbol{F^{\rq}_{t} \theta_{t}} + \nu_{t} \\ \nonumber
\boldsymbol{\theta_{t}} &=&\boldsymbol{ G_{t} \theta_{t - t}} + \boldsymbol{\omega_{t}}\nonumber
\end{eqnarray} \label{eq:3}
where
%	 \begin{eqnarray}
%    %\color{blue} y_t &=& \color{blue} \beta_0 + \beta_1 x_t + \epsilon_t \color{black} \nonumber \\ \nonumber
%    %\color{red} y_{t-1} &=& \color{red}  \beta_0 + \beta_1 x_{t-1} + \epsilon_{t-1} \color{black} 
%  \end{eqnarray}
\begin{eqnarray} %\label{eq:6}
\boldsymbol{\nu_{t} \sim N(0,V_{t})}\\ \nonumber
%\color{black} \omega_{t} &\sim& \color{black} N(0,W_{t}) \\ \nonumber
\boldsymbol{\omega_{t} \sim N(0,W_{t})} \nonumber
\end{eqnarray} \label{eq:4}

%$\boldsymbol{\theta_{t}}$

Here, $\boldsymbol{\theta_{t}}$ represent the forecast parameter $\boldsymbol{F_{t}}$ %\textbf{$F_{t}$} 
where $\boldsymbol{F_{t}}$ is a known
$n \times r$ matrix of independent variables, $\boldsymbol{G_{t}}$ is a known $n \times n$ system matrix, $\boldsymbol{W_{t}}$ is a known $n \times n$ evolution variance matrix, and $\boldsymbol{V_{t}}$ is a known $r \times r$ observational variance matrix.
%\begin{itemize}
%\item $\boldsymbol{F_{t}}$ is a known $ n \times r$ matrix of independent variables
%\item $\boldsymbol{G_{t}}$ is a known $ n \times n$ system matrix
%\item $\boldsymbol{W_{t}}$ is a known $n \times n$ evolution variance matrix 
%\item $\boldsymbol{V_{t}}$ is a known $r \times r$ observational variance matrix
%
%\end{itemize}

The prior forecast parameter $\boldsymbol{\theta_{t}}$ is found by noting $(\boldsymbol{\theta_{t-1}}|D_{t-1})\boldsymbol{\sim N(m_{t-1},C_{t-1})}$ for some mean $\boldsymbol{m_{t-1}}$ and variance matrix $\boldsymbol{C_{t-1}}$. The prior for $\boldsymbol{\theta_{t}}$  may be seen to be $(\boldsymbol{\theta_{t}}|D_{t-1}) \boldsymbol{\sim N(a_{t},R_{t})}$ whereby $\boldsymbol{a_{t} = G_{t}m_{t-1}}$ with $\boldsymbol{R_{t} = G_{t}C_{t-1}G^{\rq}_{t} + W_{t}}$. The one step ahead forecast is calculated as $(Y_{t}|D_{t-1})\sim N(f_{t},Q_{t})$. Here, $f_{t}$ is the current treatment allocation for patient $y$, while $Q_{t}$ is the forecast allocation variance for patient $y$. The posterior for $\boldsymbol{\theta_{t}}$ relies on
$(\boldsymbol{\theta_{t-1}}|D_{t-1})\boldsymbol{\sim N(m_{t},C_{t})}$

Furthermore, $\boldsymbol{m_{t} = m_{t-1} + A_{t}}e_{t}$, where $\boldsymbol{m_{t}}$ represents the current mean matrix, $\boldsymbol{C_{t} = R_{t}-A_{t}}Q_{t}\boldsymbol{A^{\rq}_{t}}$ where $\boldsymbol{C_{t}}$ is the current variance matrix, $\boldsymbol{A_{t} = R_{t}F_{t}}Q^{-1}_{t}$ where $\boldsymbol{A_{t}}$ is the adaptive coefficient, and $e_{t} = Y_{t} - f_{t}$ represents the error term.

\section*{Random Allocation Methods}
Random Allocation models of  \cite{zhang2006response} proposed the solution to minimizing the responses by using

\begin{centering}
	\begin{eqnarray} \label{eq:5} \label{eq:5} \label{eq:5} 
	%\begin{split}
	w_{A} &=&  
	\begin{cases}
	\frac{Q_{A_t}\sqrt{f_{B_t}}}{Q_{A_t}\sqrt{f_{B_t}}+Q_{B_t}\sqrt{f_{A_t}}} & \mbox{if ($f_{A_t}<f_{B_t}\mid\frac{Q_{A_t}\sqrt{f_{B_t}}}{Q_{B_t}\sqrt{f_{A_t}}}>1$)}\\% $\mid \mid$
	\frac{Q_{A_t}\sqrt{f_{B_t}}}{Q_{A_t}\sqrt{f_{B_t}}+Q_{B_t}\sqrt{f_{A_t}}} &\mbox{if ($f_{A_t}>f_{B_t}\mid\frac{Q_{A_t}\sqrt{f_{B_t}}}{Q_{B_t}\sqrt{f_{A_t}}}<1$)}  \\
	\frac{1}{2} & \mbox{Otherwise}  \\
	\end{cases}   \\ 
	%\end{split}
	%	\end{eqnarray}
	%	\begin{eqnarray} \label{eq:11}
	w_{B} &=& 1 - w_{A}  \nonumber
	\end{eqnarray} %\label{eq:12}
\end{centering}

as an optimal method to obtain weighted allocation values. However, \cite{biswas2009optimal} demonstrated the design of \cite{zhang2006response} was slightly flawed for negative values involving at least one of either $f_{A_t}$ or $f_{B_t}$. The optimal design solution posed by \cite{biswas2009optimal}  was shown to be

\begin{center}
%	\begin{figure*}
	\begin{eqnarray}  \label{eq:6}
	w_{A} &=& \frac{Q_{A_t}\sqrt{\gamma_{B_t}}}{Q_{A_t}\sqrt{\gamma_{B_t}}+Q_{B_t}\sqrt{\gamma_{A_t}}}  \\ \nonumber 
	w_{B}&=&1-w_{A}\\ \nonumber
	\textnormal{where}\
	 \gamma_{A}&=&\Phi\left(\frac{f_{A_t}-f_{B_t}}{\sqrt{Q^2_{A_t}+Q^2_{B_t}}}\right),\\
	 \gamma_{B}&=&\Phi\left(\frac{f_{B_t}-f_{A_t}}{\sqrt{Q^2_{A_t}+Q^2_{B_t}}}\right)  \nonumber
	\end{eqnarray}  \nonumber
	%\end{figure*}
\end{center}
Recently, \cite{donahue2020allocation} examined how a Decreasingly Informative Prior distribution impacted the allocation using each of these equations. The current work uses the DLM to randomly allocate patients to examine these impacts. %A covariate is then added and a comparison made. 
Yet because the DLM is an updating method at each value, the values for each of $f_{A_t}, f_{B_t}, Q_{A_t}, Q_{B_t}$ will change at each iteration, leading to different weight values based on the starting values.

\section*{Algorithm}
To generate the allocation values
\begin{enumerate}
	
	\item Initiate the DLM by selecting initial values for $\mu_{A}$, $\mu_{B}$, $\omega_{t}$, $C_{t_{A}}$, $C_{t_{B}}$, $Q_{t_A}$, $Q_{t_B}$.
	\item Calculate predicted values and variances $f_{A_t}$ ($F_t = [1,0]$), $f_{B_t}$ ($F_t = [1,1]$), $Q_{A_t}$ and $Q_{B_t}$
	\item Compute $w_A$ and $w_B$
	\item Sample a Uniform(0,1) random variable U and compare $w_A$
	\item If $w_A < U$, allocate to Treatment A ($F_t = [1,0]$), otherwise allocate to treatment B ($F_t = [1,1]$)
	\item Conduct experiment and observe $y_{t}$
	\item Update the DLM and return to step 2 
\end{enumerate}

\section*{Simulation Study}

Seven scenarios were examined in \cite{donahue2020allocation} and these values may be observed in Table~\ref{tbl:SimulationScenarios}. Simulation sizes of 1,000 and 10,000 were considered and run for several scenarios however the results were almost identical, therefore, in order to avoid any unnecessary computation time the DLM was used to randomly allocate each scenario through 1000 simulations. As in \cite{donahue2020allocation} treatment allocation probabilities, total number of allocations in each treatment group, and total number of successes was recorded, however, the current authors have only included the treatment allocation associated with the preferred treatment and these may be seen in Table~\ref{tbl:TreatmentGroupMSSnocov}. Although \cite{donahue2020allocation} utilized Bayesian updating to obtain the values of the decreasingly Informative Prior, each iteration was manually done, leading to a large completion time due to the extensive number of necessary simulation calculation runs. % prior to a two sample t-test to determine when the allocation algorithm switched treatment allocations using a $p-value = 0.05$ in a one tailed test. 
In the current method using the DLM, these times were greatly reduced. Each scenario was run using R Studio version 1.2.1335 on an ACER computer with an AMD Ryzen 5 2500U with Radeon Vega Mobile Gfx 2.00 GHz processor and 8.00 GB of RAM using Windows 10. Additionally, each run took approximately 45 seconds to complete, with the longest run time 164 seconds corresponding to the budget size $N = 200$, while the shortest run time 23 seconds corresponding to a budget size $\mathscr{N} = 34$.  
\begin{table}[ht]
		\small\sf\centering
	\begin{center} \captionsetup{justification=centering} \caption{Simulation Scenarios}
	\begin{adjustbox}{max width=\columnwidth}
		\begin{tabular}{c c c c c c c c c | r r r r r r r r r } 
		\toprule	
			Scenario & Differences & Standard & Planned Sample\\ 
			$$&$$&Deviation&Budget \\ 
			\midrule
			1 &	0 &		20 &	128\\
			2 &	10 &		15 &	74\\
			3 &	10 &	    20 &	128\\
			4 &	10 &		25 &	200\\
			5 &	20 &	 	20 &	34\\
			6 &	20 &		25 &	52\\
			7 &	20 &		30 &	74\\
\bottomrule
		\end{tabular}\label{tbl:SimulationScenarios}
	\end{adjustbox}
	\end{center}
\end{table}

The results for a mean difference of 0 and standard deviation of 20 may be seen in Table~\ref{tbl:TreatmentGroupMSSnocov} and a plot of both equal and unequal allocation may be observed in Figure~\ref{fig:AllocationFirstFormulaNoCovariate} below. The mean number of allocations was obtained using each method. Notice the mean allocation using the method of \cite{zhang2006response} was 63.538, which is as expected, given the probability of allocation to Treatment A was 0.5. One may observe this outcome in  Figure~\ref{fig:AllocationFirstFormulaNoCovariate}a where no allocation differences exist .
\begin{table}[ht]
		\small\sf\centering
	\begin{center}\caption{Treatment Group Mean Sample Size. Italicized values indicate Treatment B was selected}
		\begin{adjustbox}{max width=\columnwidth}
		\begin{tabular}{c c c c c c }
			\toprule
			Mean & Standard&Sample&Equation \ref{eq:5}& Equation \ref{eq:6} \\
			Difference&Deviation&Budget&Allocation&Allocation \\
			\midrule
			0 &	20 &		128 &	$63.538$&$\textit{32.745}$\\
			10 &	15 &		74 &	$36.470$&$\textit{4.928}$\\
			10 &	20 &	    128 &	$63.146$&$\textit{13.283}$\\
			10 &	25 &		200 &	$\textit{99.883}$&$\textit{34.589}$\\
			20 &	20 &	 	34 &	$16.673$&$\textit{2.390}$\\
			20 &	25 &		52 &	$25.641$&$\textit{2.793}$\\
			20 &	30 &		74 &	$\textit{36.958}$&$\textit{3.859}$\\						
		\bottomrule
			\end{tabular}\label{tbl:TreatmentGroupMSSnocov}
			\end{adjustbox}
			\end{center}
			\end{table}

When the DLM was applied to the unequal method of \cite{biswas2009optimal}, the mean number applied to Treatment A is 95.255, while the mean number allocated to Treatment B is 32.745. Under the methods of \cite{biswas2009optimal}, \cite{zhang2006response}, and \cite{donahue2020allocation}, the smaller value was taken to be the better allocation, therefore, it appears as though Treatment B is the favorable treatment.

Figure 1 Here titled AllocationFirstFormulaNoCovariate
			
			\begin{figure}[ht]
					\begin{adjustbox}{max width=\columnwidth}
				\centering
				\begin{tabular}{cc}
					\includegraphics[width=.5\textwidth]{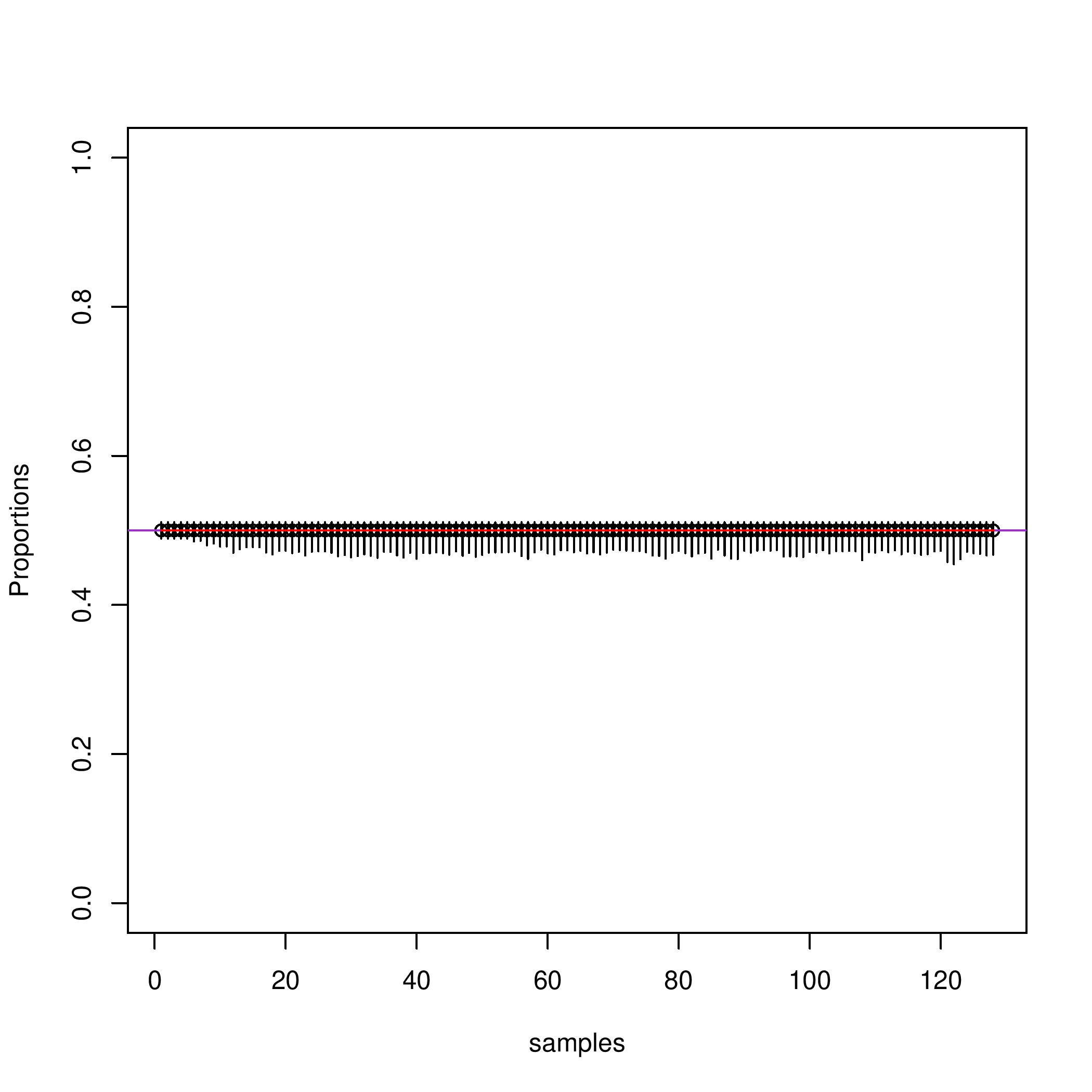} & \includegraphics[width=.5\textwidth]{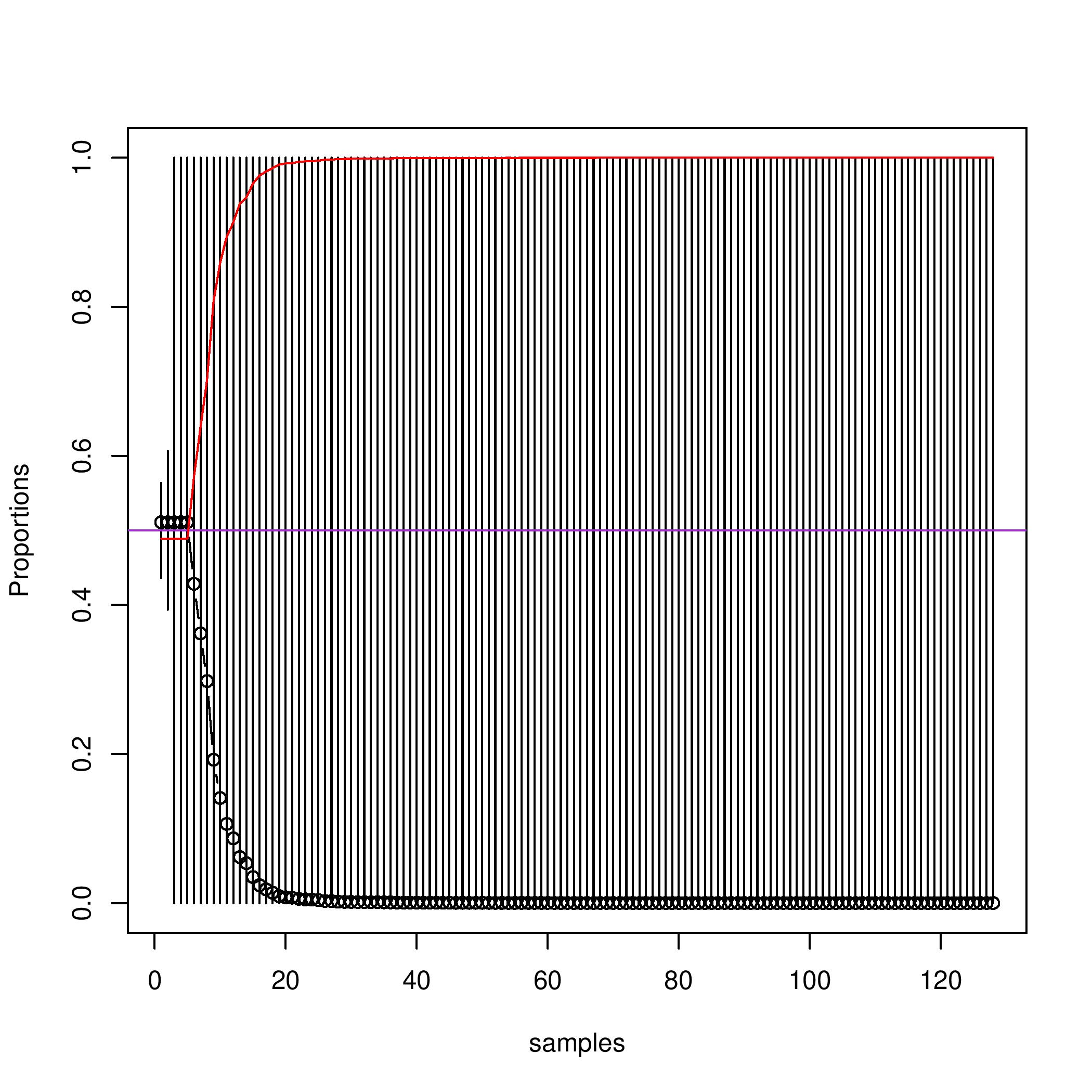}\\
					a). Equal Allocation& b). Unequal Allocation  \\
				\end{tabular}
			\end{adjustbox}
				\caption{Comparison Between Equal and Unequal Allocation} \label{fig:AllocationFirstFormulaNoCovariate}
			\end{figure}
			
An examination of Figure~\ref{fig:AllocationFirstFormulaNoCovariate}b illustrates the allocation probabilities for both Treatment A and Treatment B. Each allocation begins at 0.5, however, dependent upon the particular treatment which was allocated, the weights either increase or decrease. The mean allocation weight for treatment A was 0.749, while the mean weight for Treatment B was 0.251. %Both allocations  had a deviation value of 0.396. 
The weighted values for Treatment A are seen in Figure~\ref{fig:AllocationFirstFormulaNoCovariate}b as the red line, while those for Treatment B are noticeably the opposite. This is due to the symmetry between the two weighting schemes. Problematic to these two methods was the fact that with equal variance, the treatment allocation weights remained at approximately 0.5 in the method of \cite{zhang2006response}, while using the method proposed by \cite{biswas2009optimal}, the treatment allocation proportions immediately converged.
			
However, determining behavior of the treatment allocation weights upon varying the parameter values associated with the mean, system variance and observational variance values is important in determining model behavior. By analyzing model behavior through these parameter modifications clinical trial researchers can determine the minimum number of subjects necessary to detect the favorable treatment, enabling them to conclude the study earlier, thereby avoiding the ethical issues presented by the continuation of providing unfavorable treatments.
			
\begin{figure*}[ht]
\centering
\begin{adjustbox}{max width=\columnwidth}
\begin{tabular}{lccc}
\includegraphics[width=.30\textwidth]{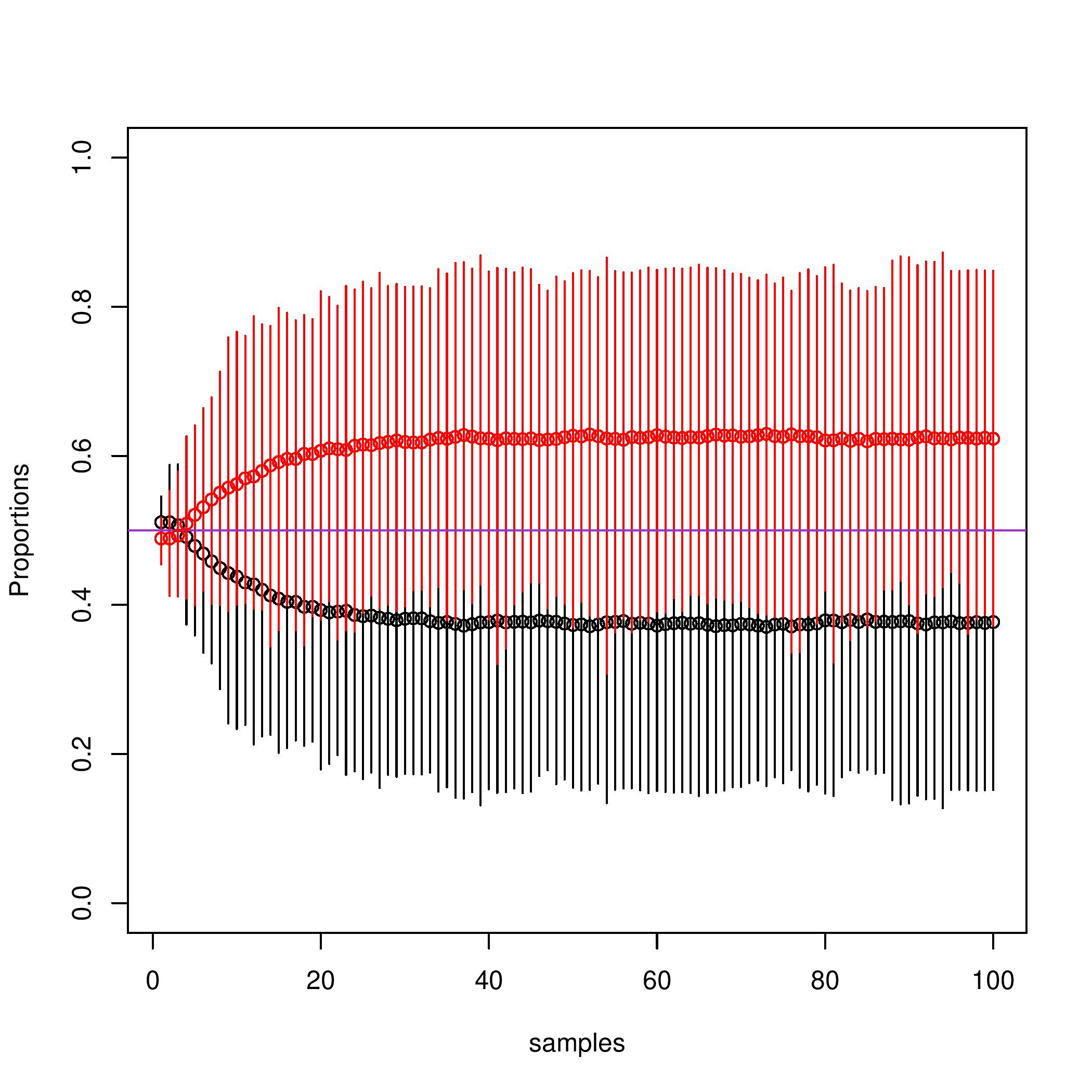} &
\includegraphics[width=.30\textwidth]{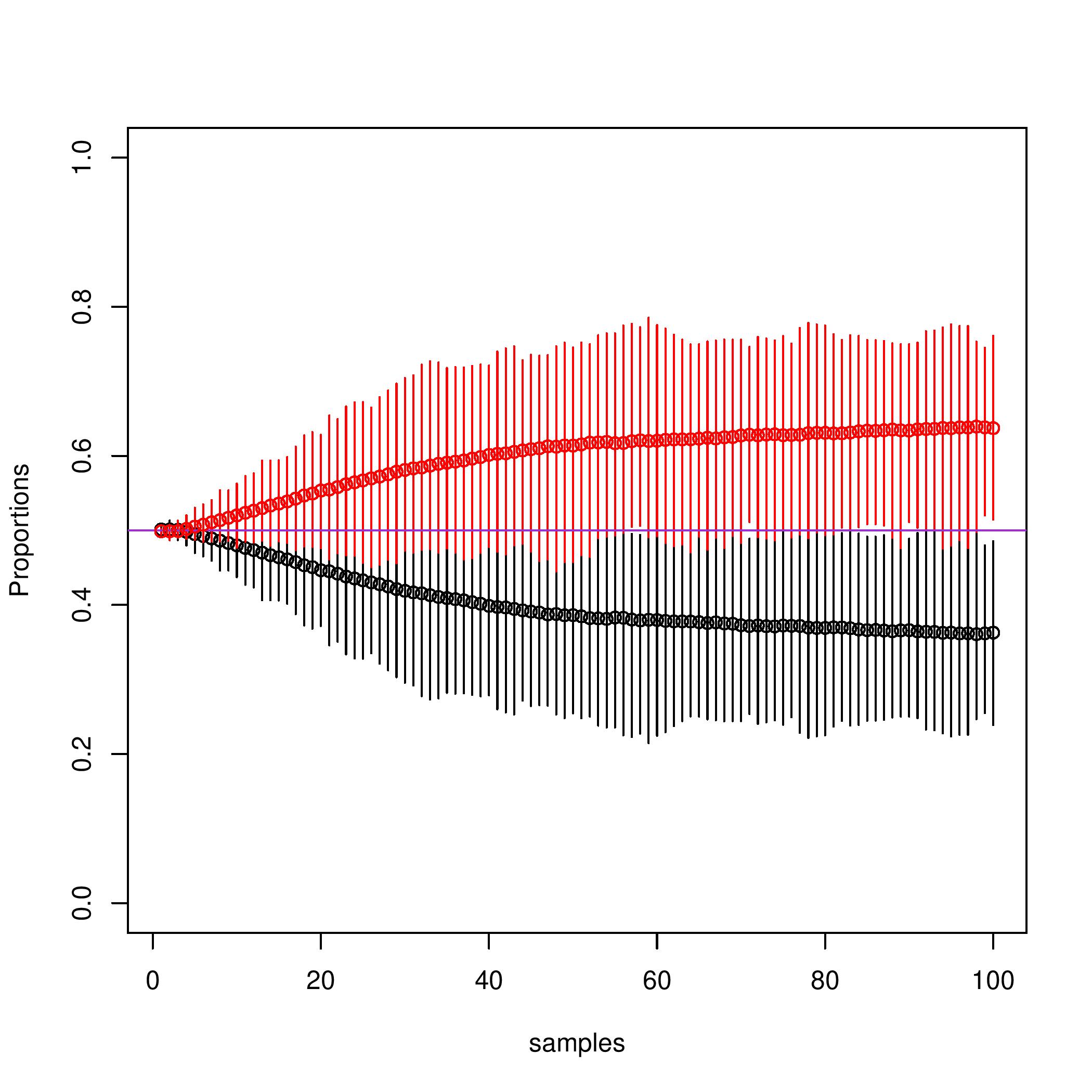}&
\includegraphics[width=.30\textwidth]{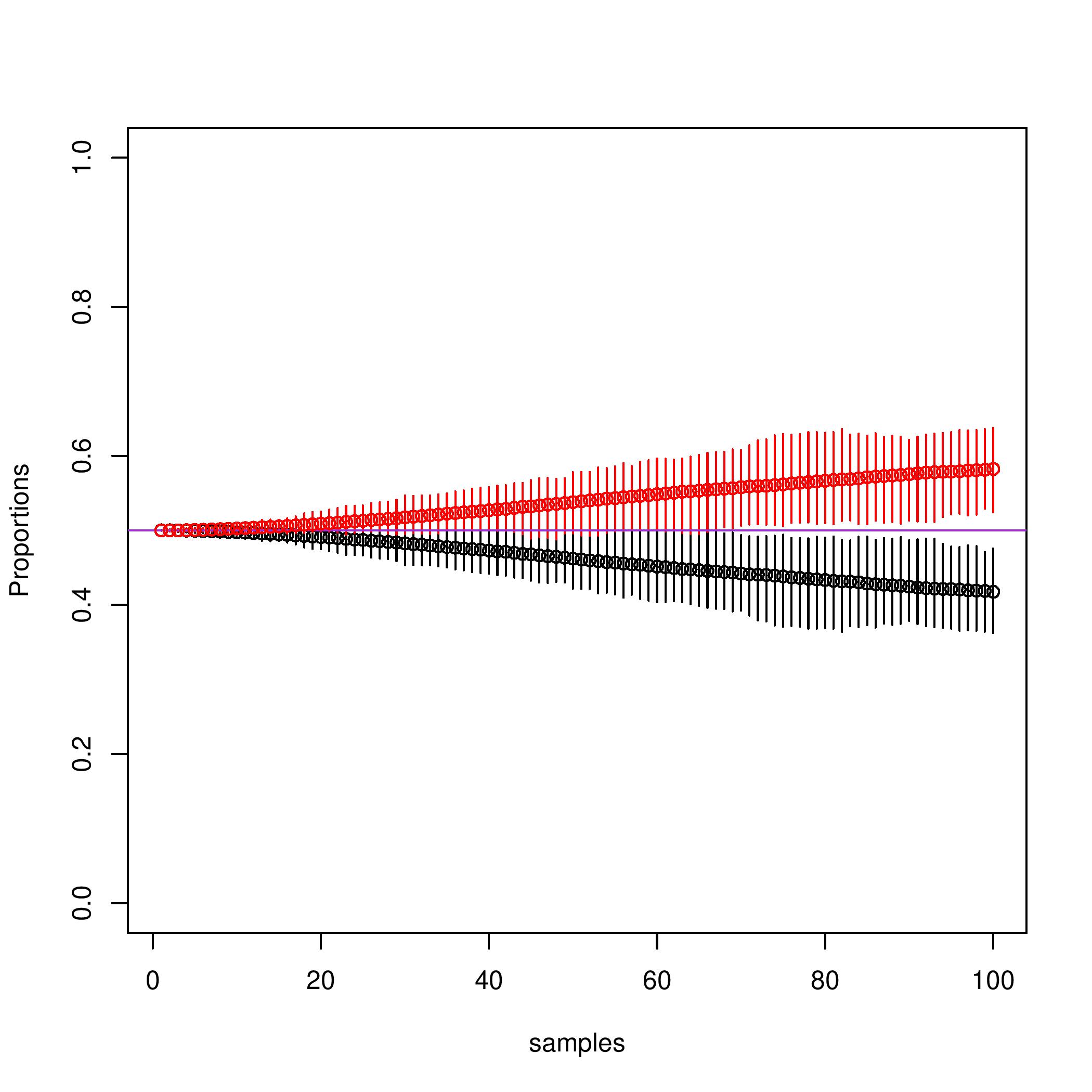}\\
(a) $\mu_{B}=1, \omega_t = 0.1$	  & (b) $\mu_{B}=1, \omega_t = 0.01$ & (c) $\mu_{B}=1, \omega_t = 0.001$\cr
\includegraphics[width=.30\textwidth]{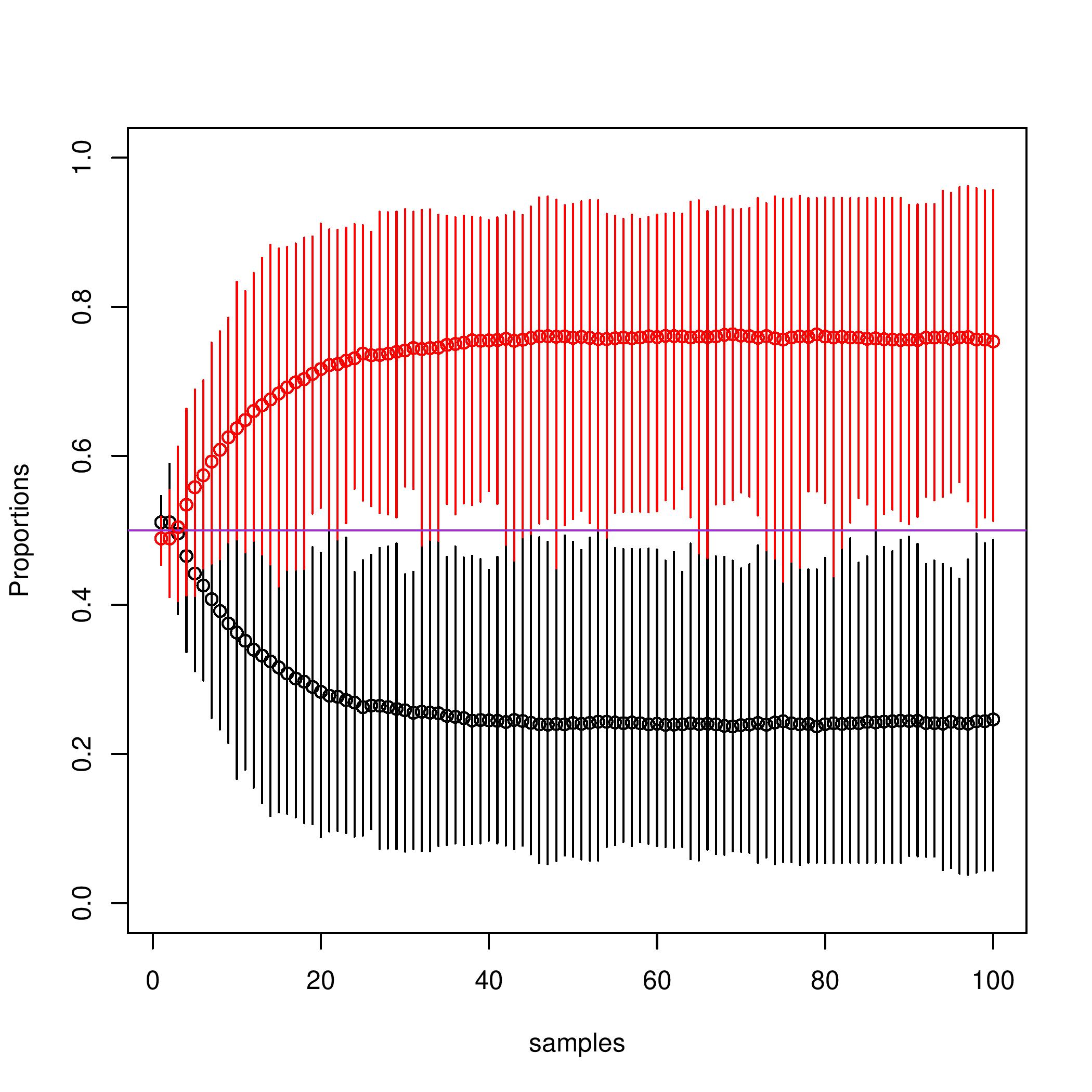} &
\includegraphics[width=.30\textwidth]{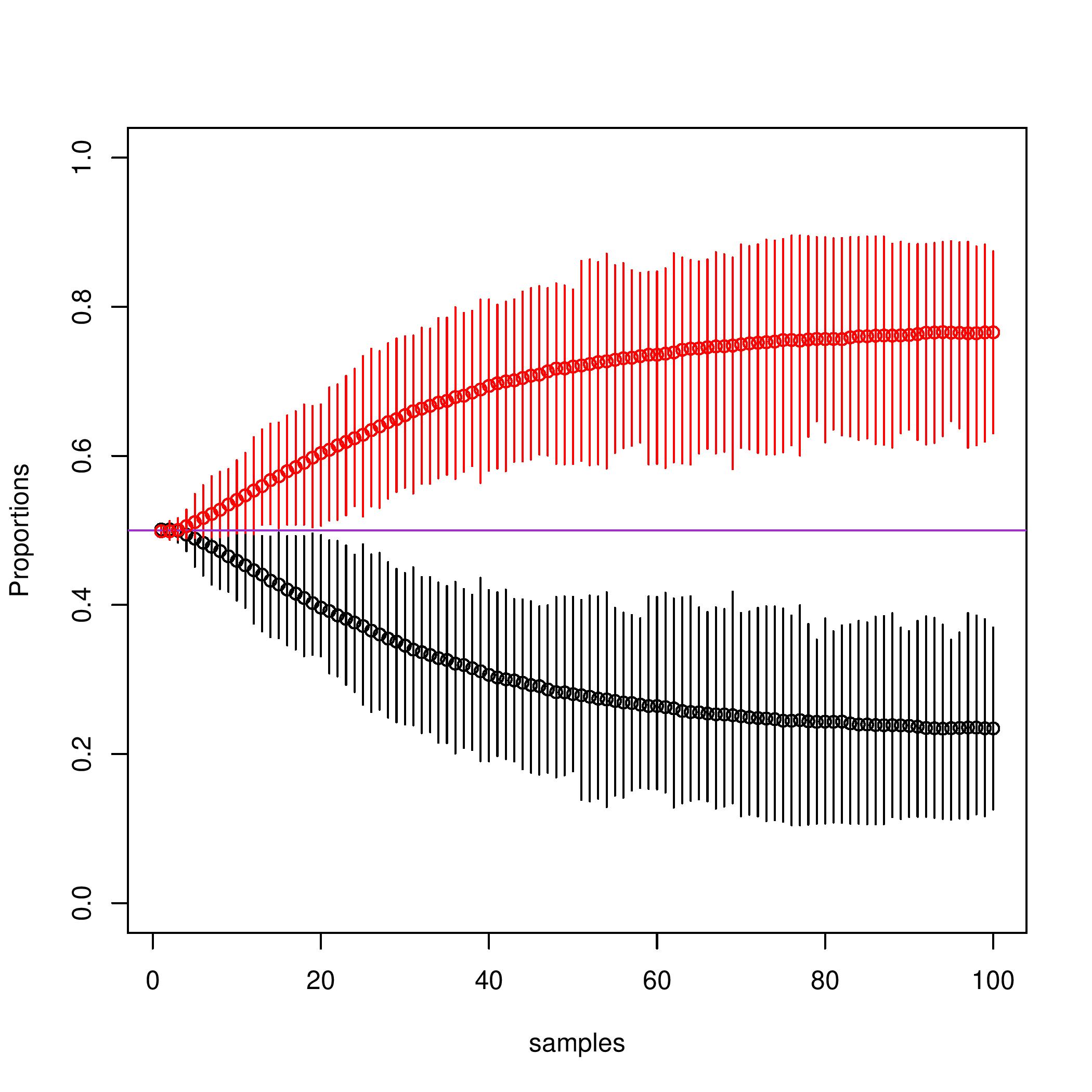}&
\includegraphics[width=.30\textwidth]{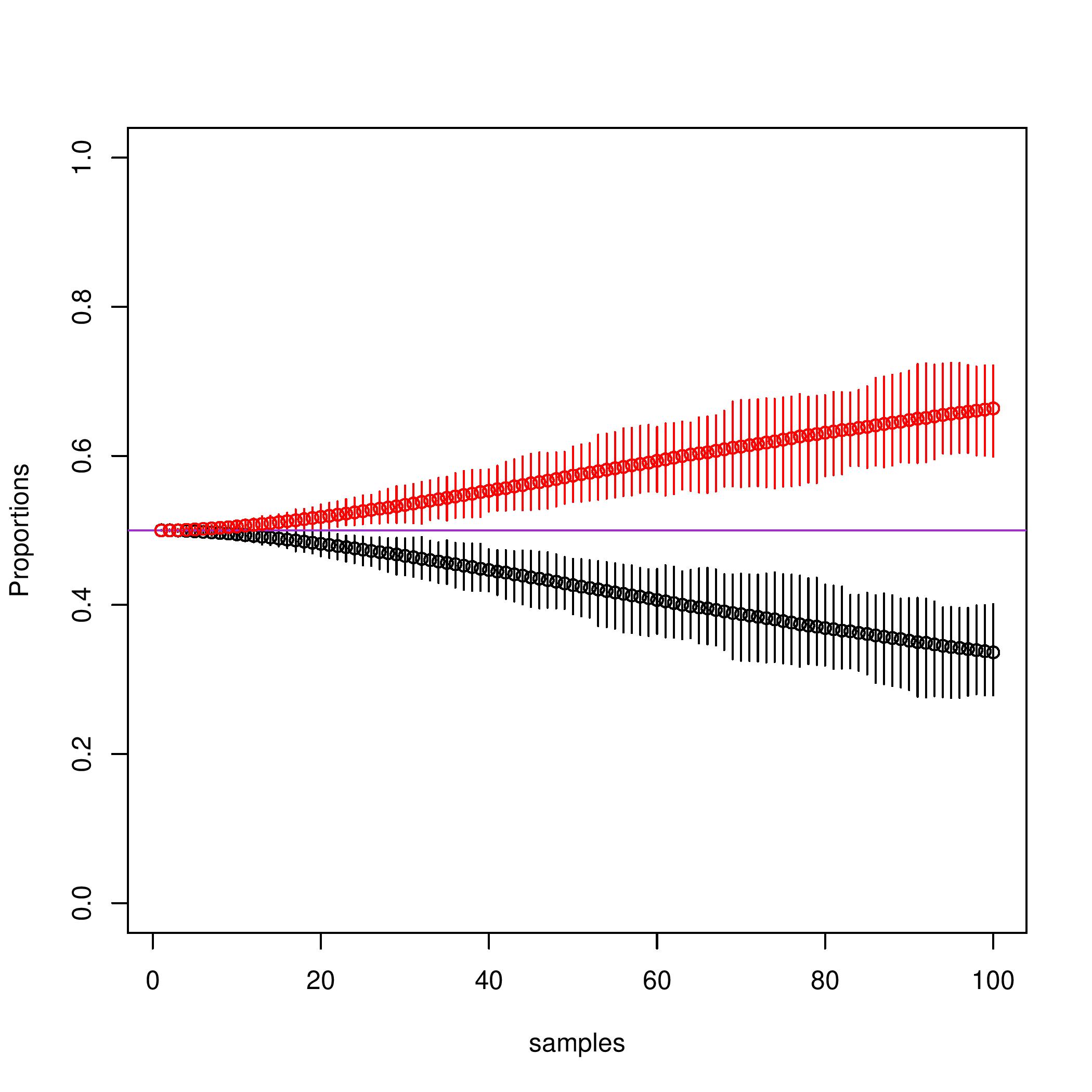}\\
(d) $\mu_{B}=2, \omega_t = 0.1$ & (e) $\mu_{B}=2, \omega_t = 0.01$ & (f) $\mu_{B}=2, \omega_t = 0.001$ \cr
\includegraphics[width=.30\textwidth]{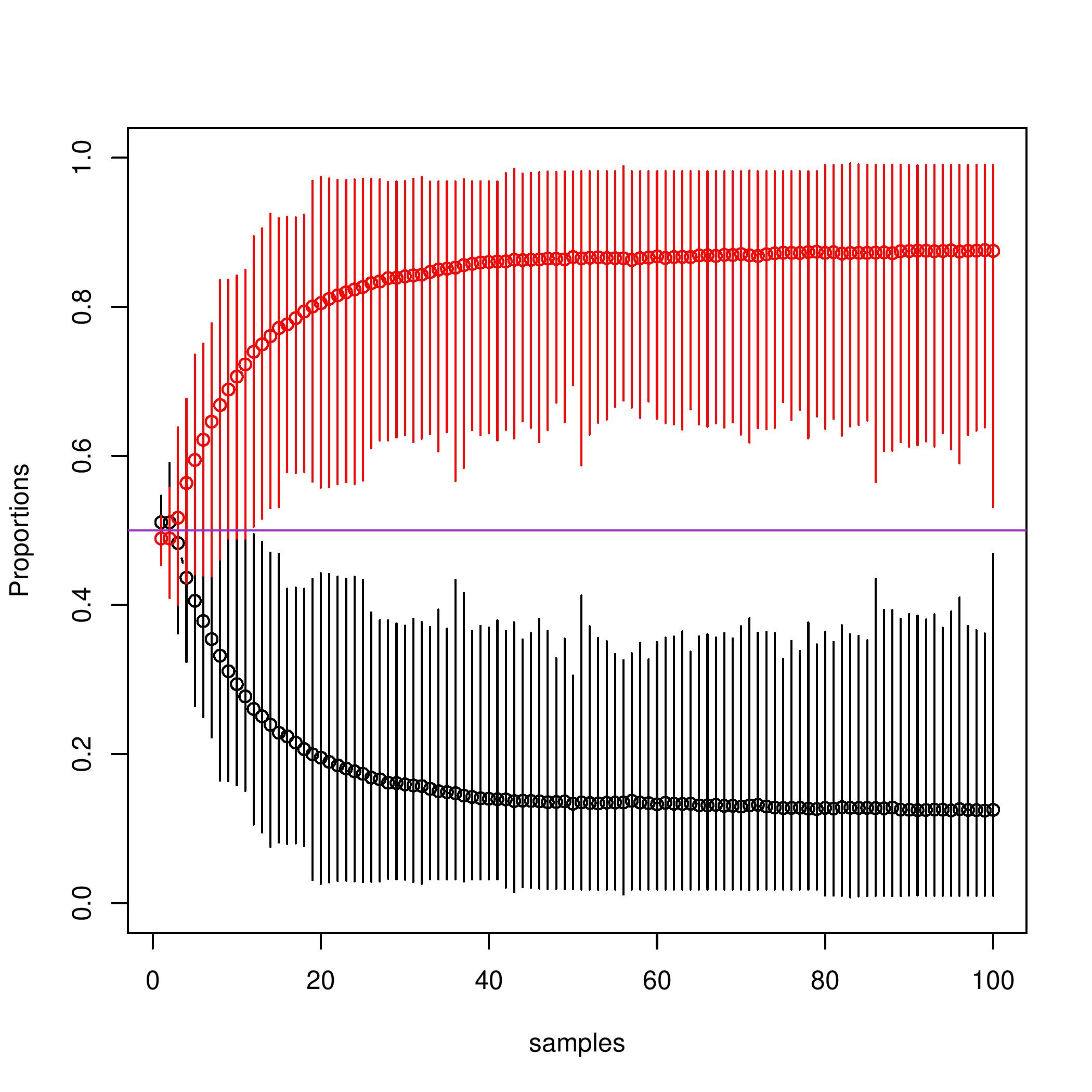} &
\includegraphics[width=.30\textwidth]{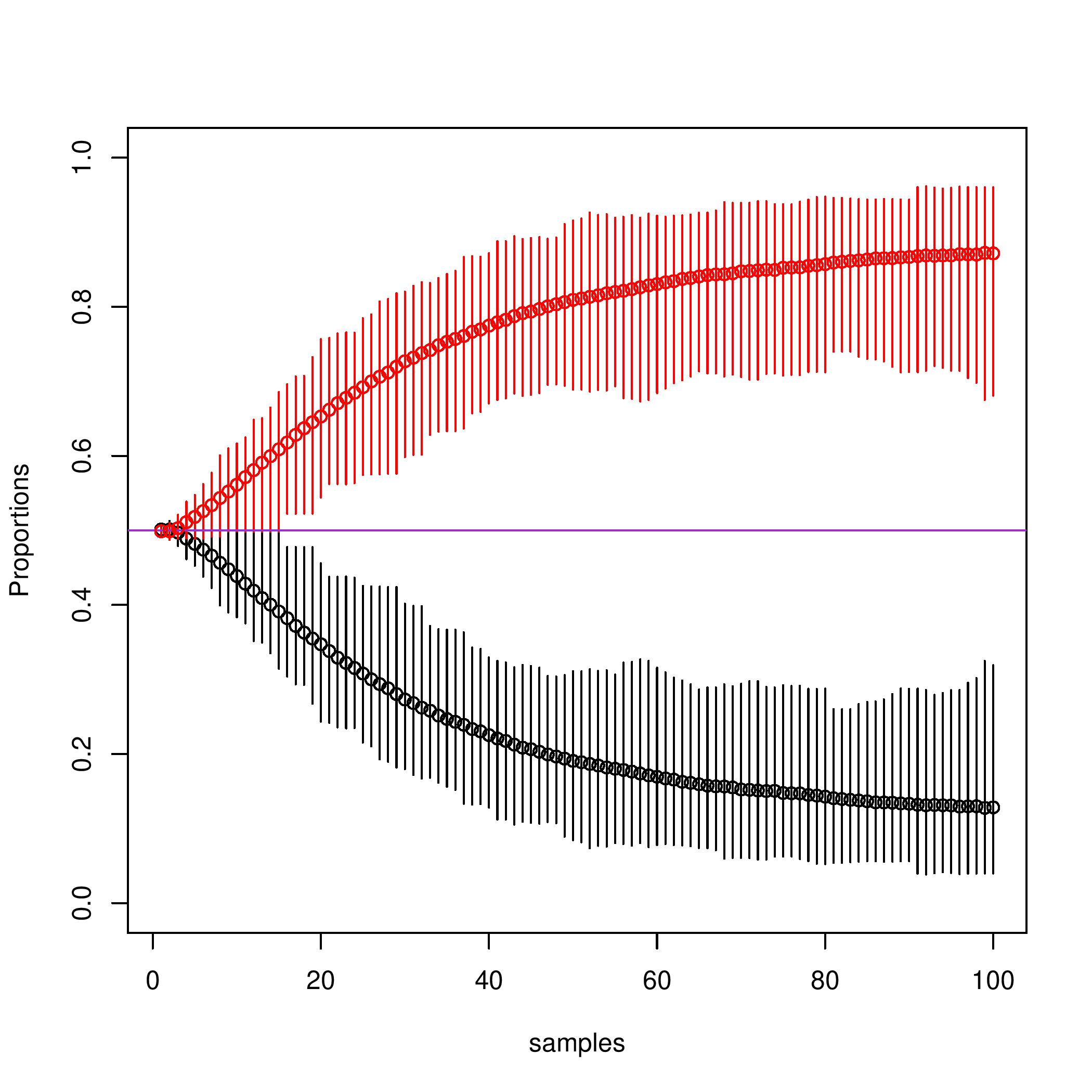}&
\includegraphics[width=.30\textwidth]{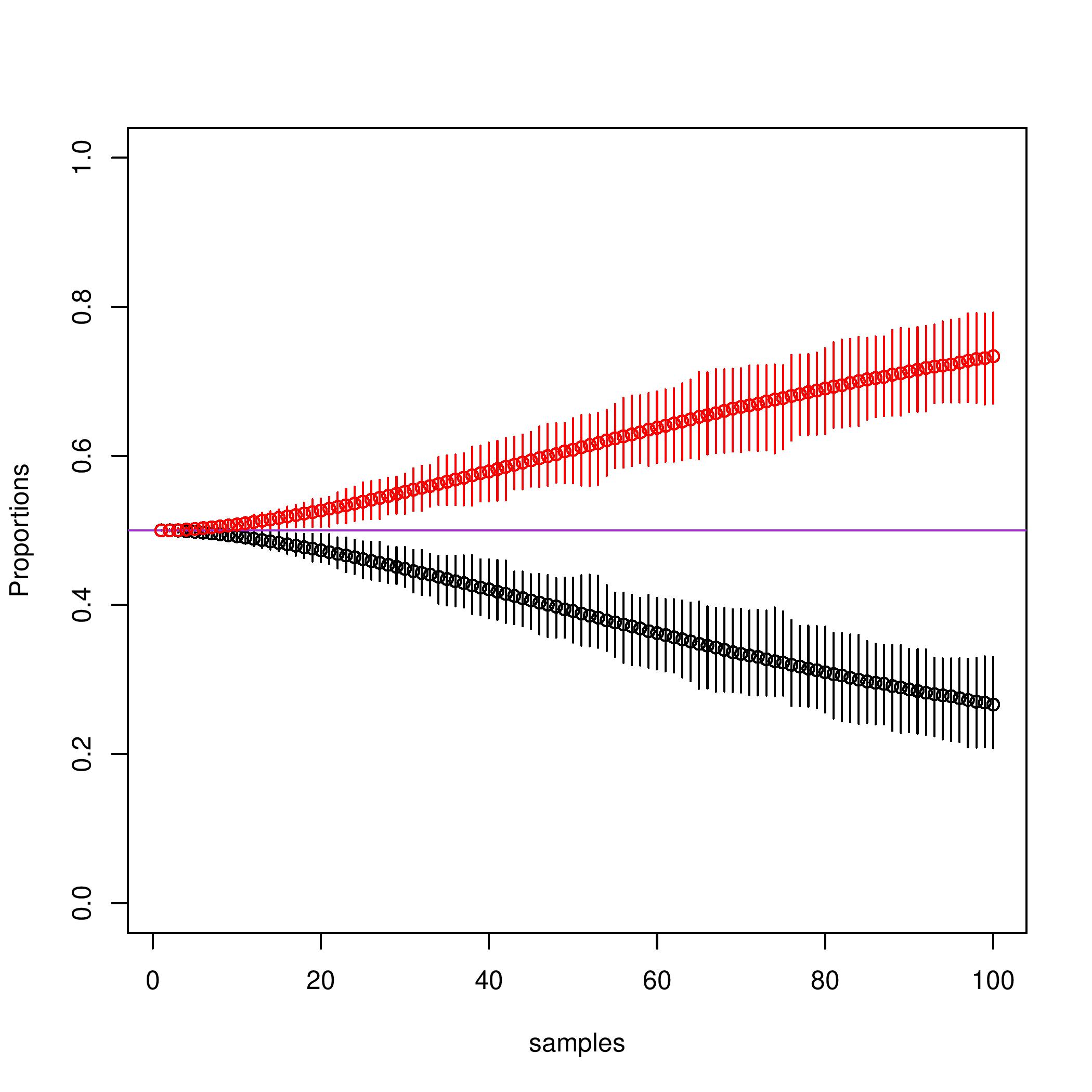}\\
(g) $\mu_{B}=3, \omega_t = 0.1$& (h) $\mu_{B}=3, \omega_t = 0.01$ & (i) $\mu_{B}=3, \omega_t = 0.001$ \cr
\includegraphics[width=.30\textwidth]{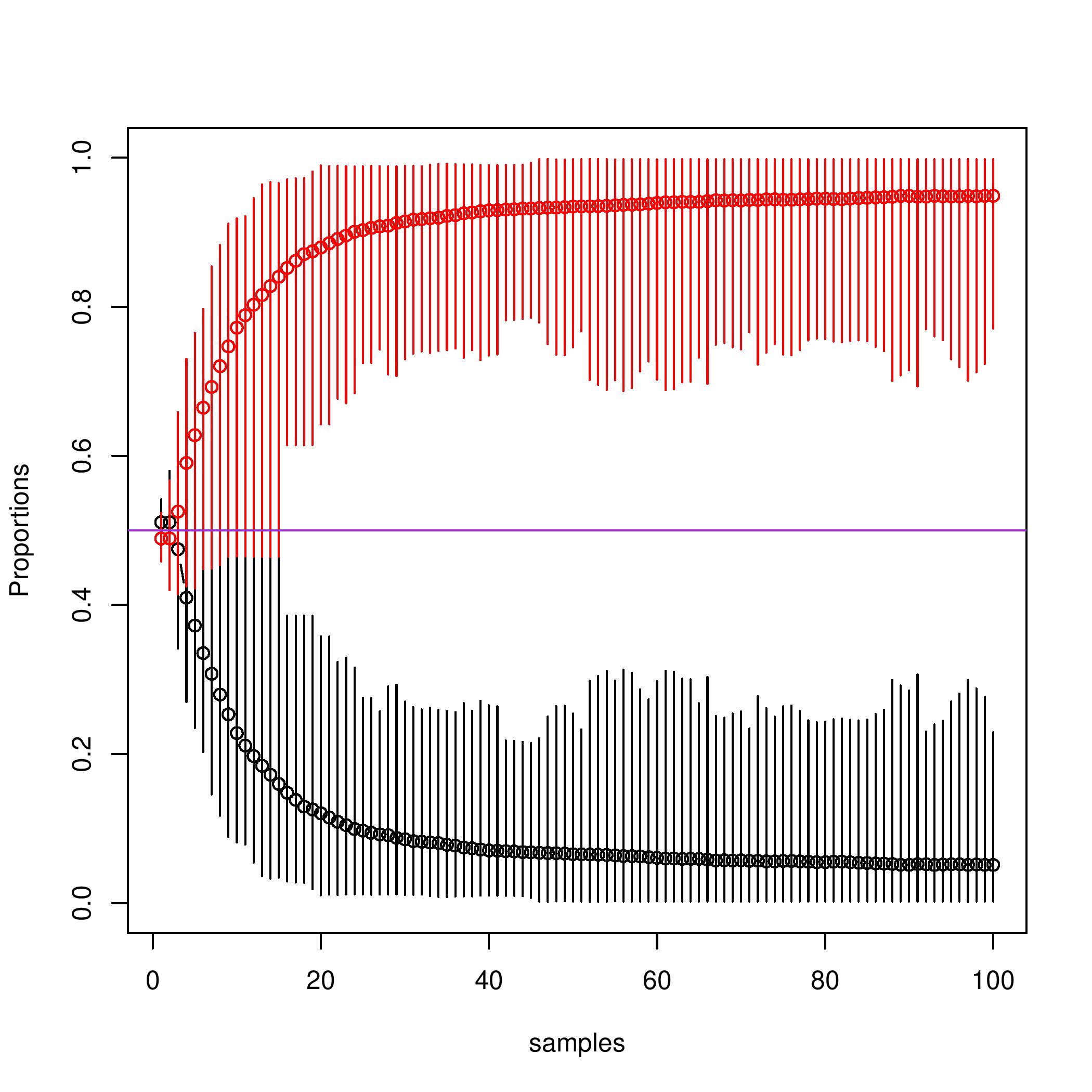} &
\includegraphics[width=.30\textwidth]{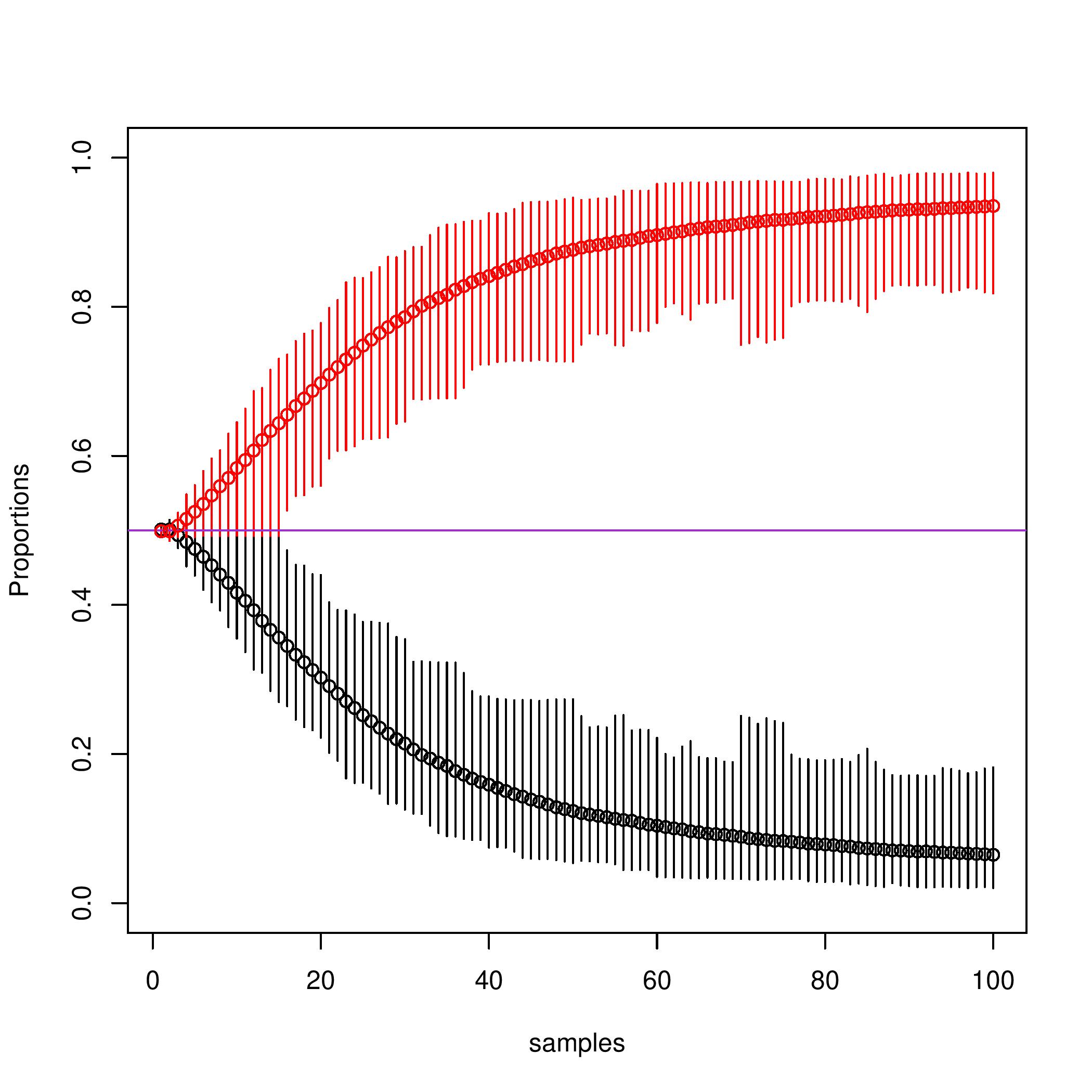}&
\includegraphics[width=.30\textwidth]{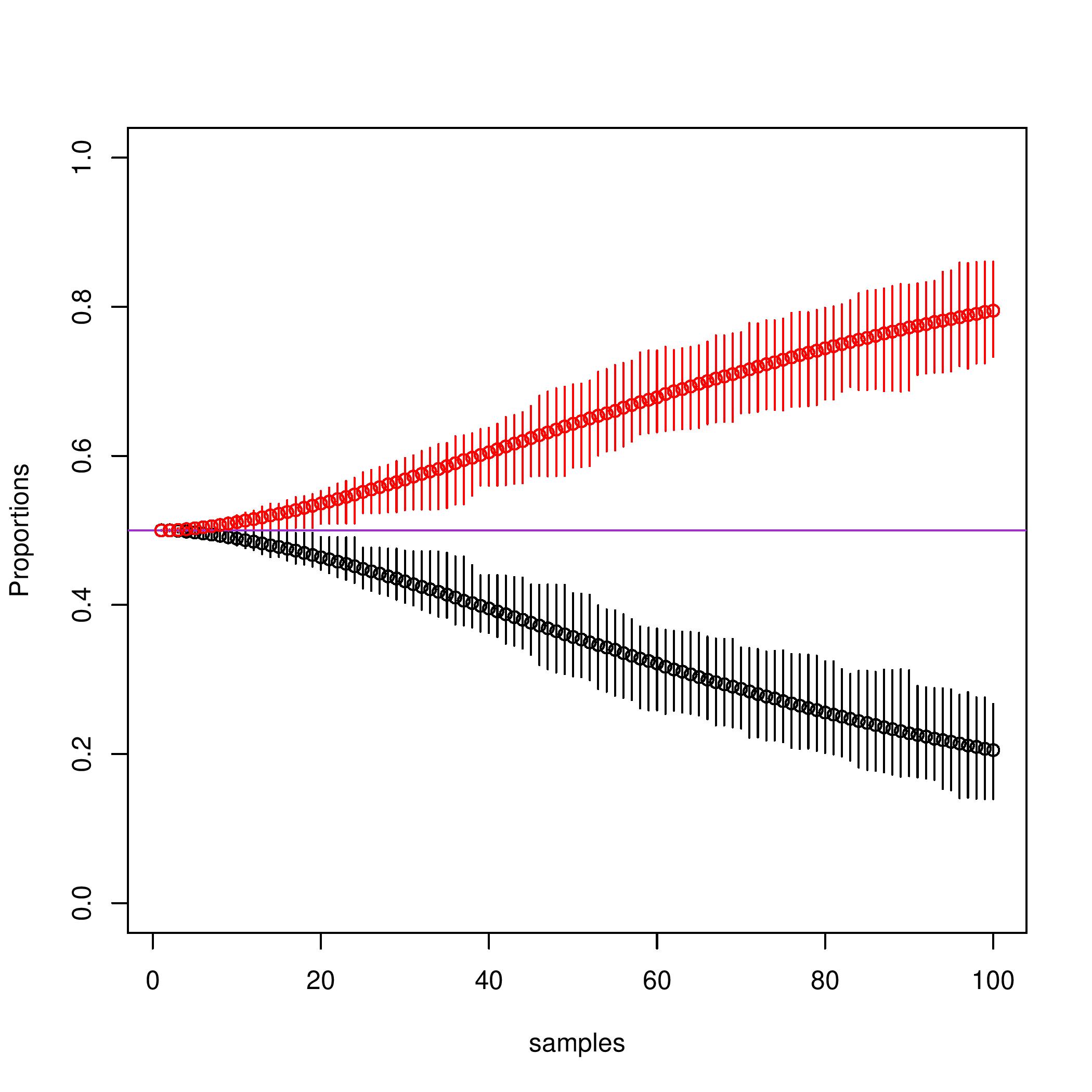}\\
(j) $\mu_{B}=4, \omega_t = 0.1$ & (k) $\mu_{B}=4, \omega_t = 0.01$ & (l) $\mu_{B}=4, \omega_t = 0.001$ \cr
\includegraphics[width=.30\textwidth]{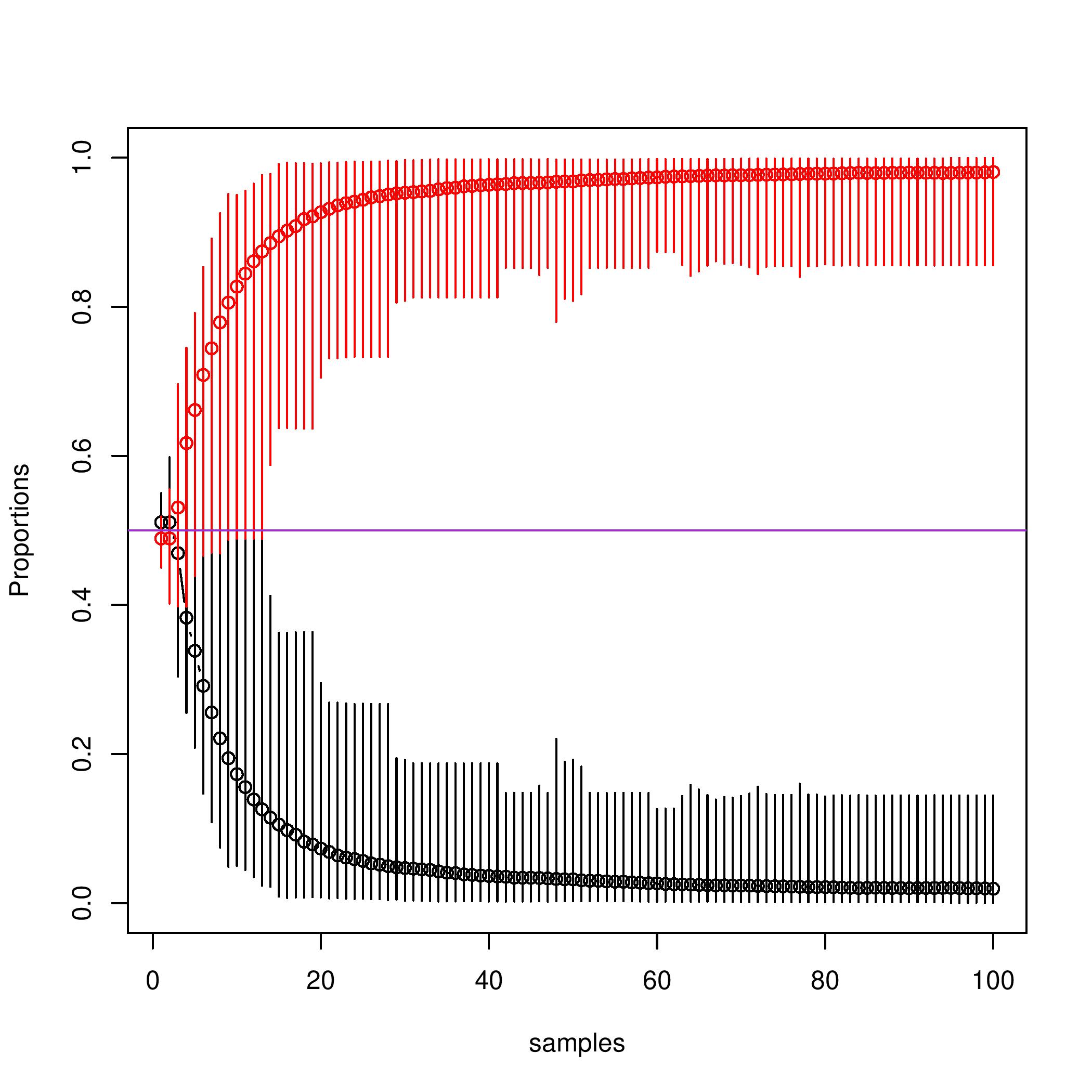} &
\includegraphics[width=.30\textwidth]{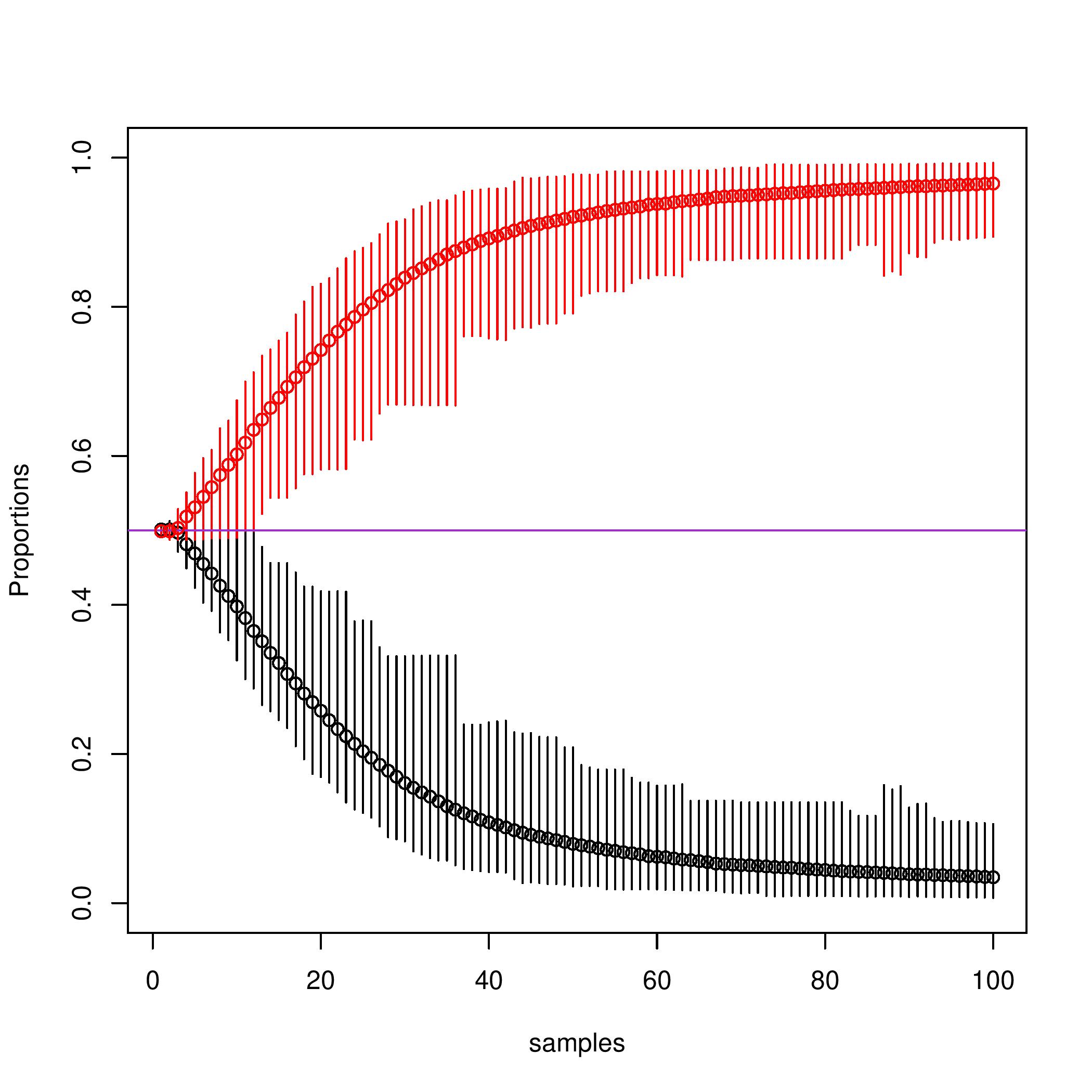}&
\includegraphics[width=.30\textwidth]{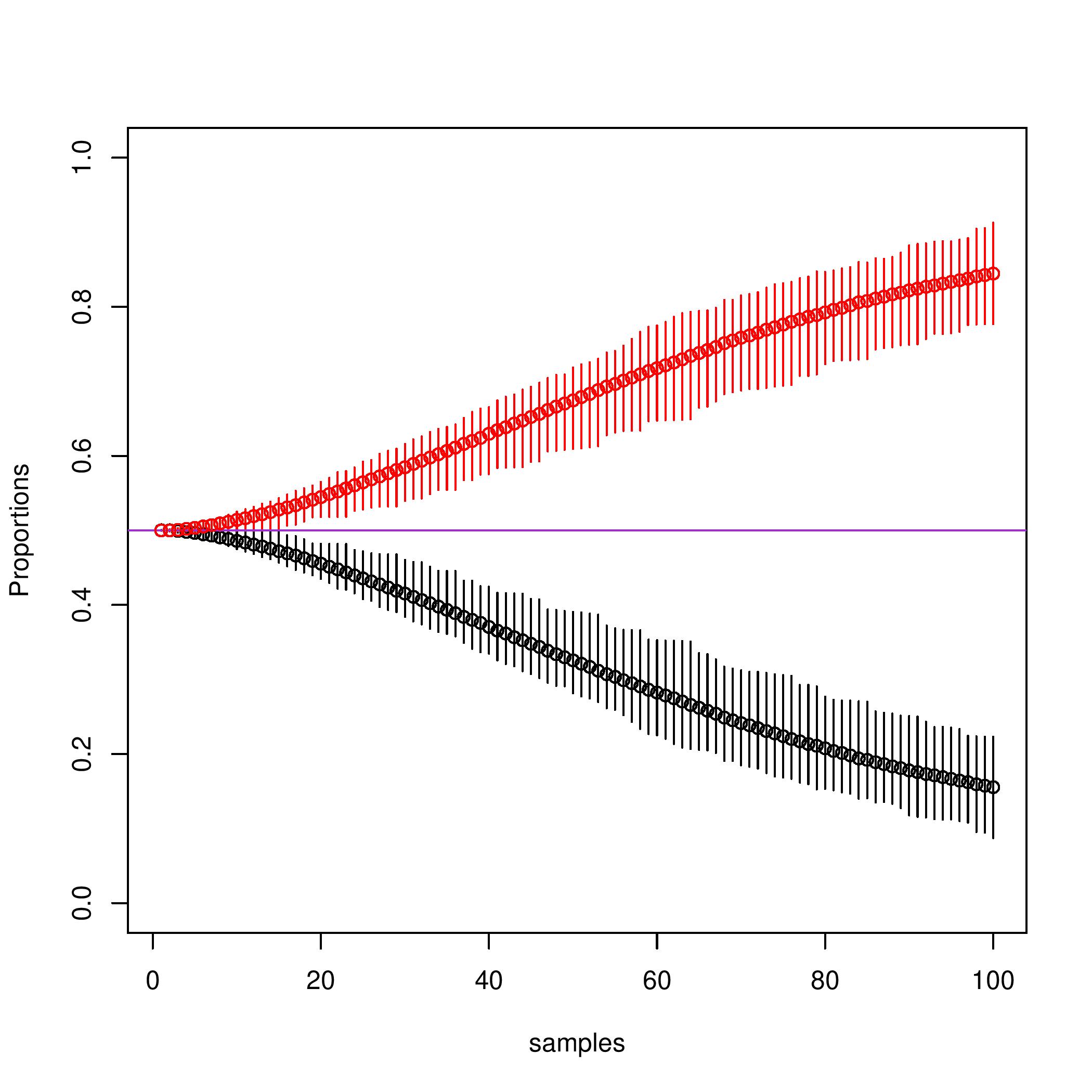} \\
(m) $\mu_{B}=5, \omega_t = 0.1$ & (n) $\mu_{B}=5, \omega_t = 0.01$ & (o) $\mu_{B}=5, \omega_t = 0.001$ \\
\end{tabular}
\end{adjustbox}
\caption{Comparison of Weight Allocation proportions for $\omega_{t} = 0.1, 0.01$ and $0.001$ and $\mu_{B} = 1, 2, 3, 4, 5$ and $C_{t_B} =0.000001$ with bars representing the uncertainty across simulations.} 
\label{fig:compweightA}
\end{figure*}

Therefore, the current authors chose a budget size of 100 and a sensitivity analysis was conducted using various values for $\mu_{B}$, $\omega_t$, and $c_{t_B}$, while keeping $Q_t = 1$. The values chosen for $\mu_B$ were 1 - 5, leading to $H_{A}: \mu_{B} = 1$ through $H_{A}: \mu_{B} = 5$. This lead to the hypothesis
\begin{center}
	\begin{eqnarray}  
	H_{0}: \mu_{B} &=& 0 \cr
	H_{A}: \mu_{B} &\ne& 0 
	\end{eqnarray} \label{eq:7}
\end{center}
whereby $\mu_{B} = {1, 2, 3, 4, 5}$
By keeping $Q_t = 1$, and using the patient budge size of 100, the values chosen for $\mu_{B}$ represented a 1\% to a 5\% difference in the two treatments. The values for $\omega_t$ were chosen as 0.1, 0.01, and 0.001, which represent decreased variability between times, thereby increasing certainty of between time variability impact. Finally, the values for $c_{t_B}$ were chosen to be 0.1, 0.001, and 0.000001. These values were chosen to represent an increased knowledge group B has no effect. Some of these weighted allocation proportion values may be observed in Figure~\ref{fig:compweightA}. It must be noted these were not all the weighted allocation proportion values, and these represent each of the $\mu_B$ values chosen, and each of the $\omega_t$ values chosen, but only the $c_{t_B} = 0.000001$ to illustrate the impact. 
Using a mean $\mu_B = 1$ with $\omega_t = 0.1$ the mean proportion of allocation values to treatment A was 0.607, while the mean proportion allocated to treatment B was 0.393, which may be observed in Figure~\ref{fig:compweightA}a. Furthermore, the mean number at which the treatment allocation switched from B to A was 39.749. Compare this to the treatment proportions when $\omega_t = 0.01$ in Figure~\ref{fig:compweightA}b. Here, the mean proportion of allocation values to treatment A was 0.595, while the mean proportion allocated to treatment B was 0.405. Likewise, the mean number at which the treatment allocation switched from B to A was 41.281. Finally, letting $\omega_t = 0.001$ one may observe in Figure~\ref{fig:compweightA}c
the mean proportion of allocation values to treatment A was 0.538, while the mean proportion allocated to treatment B was 0.462, with the mean number at which the treatment allocation switched from B to A was 46.730.
			
Next the mean was increased to 3, $\mu_B = 3$ and the analysis was conducted. When using $\omega_t = 0.1$ the mean proportion of allocation values to treatment A was 0.796, while the mean proportion allocated to treatment B was 0.204, which may be observed in Figure~\ref{fig:compweightA}g. Interestingly, the mean number at which the treatment allocation switched from B to A decreased from 37.749 using $\mu_B = 1$ to 18.156 using $\mu_B = 3$. When $\omega_t = 0.01$ one may see in Figure~\ref{fig:compweightA}h the mean proportion of allocation values to treatment A was 0.753, while the mean proportion allocated to treatment B was 0.247. This led to a the mean number necessary to switch from treatment B to treatment A to decrease from  41.281 at $\mu_B = 1$ to 24.450 using $\mu_B = 3$. Lastly, when $\omega_t = 0.001$ the mean proportion of allocation values to treatment A was 0.610, while the mean proportion allocated to treatment B was 0.390, which may be observed in Figure~\ref{fig:compweightA}i. Once again the mean number at which the treatment allocation switched from B to A decreased from 46.730 using $\mu_B = 1$ to 39.185 using $\mu_B = 3$, however, this value is slightly higher than when using $\omega_t = 0.01$. 
			
Finally, the output was analyzed when $\mu_B = 5$. When using $\omega_t = 0.1$ the mean proportion of allocation values to treatment A was 0.892, while the mean proportion allocated to treatment B was 0.108, which may be observed in Figure~\ref{fig:compweightA}m. The mean number at which treatment allocation went from B to A was 8.052, which is much lower that the mean values for $\omega_t = 0.1$ when using $\mu_B=$ 1 or 3. When $\omega_t$ was decreased to 0.01, the mean proportion of allocation values to treatment A was 0.832, while the mean proportion allocated to treatment B was 0.168, which may be observed in Figure~\ref{fig:compweightA}n. When using $\omega_t = 0.01$, the mean number at which treatment allocation switched from A to B increased from 8.052 to 15.209, which represents approximately twice the needed patient budget. Lastly, when the value for  $\omega_t$ was decreased to 0.001 the mean proportion of allocation values to treatment A was 0.669, while the mean proportion allocated to treatment B was 0.331, which may be observed in Figure~\ref{fig:compweightA}o. However, here the mean number at which treatment allocation switched from B to A increased from 15.209 to 33.538. This represents not only more than double the patient budget needed when going from $\omega_t = 0.01$ to $\omega_t = 0.001$, but a 4 times increase when going from $\omega_t = 0.1$ to $\omega_t = 0.001$
			
It appears clear that as the mean value for treatment B $\mu_B$ increases, the mean allocation probabilities also increase to higher convergent values. Likewise, the mean number of allocations necessary to switch from treatment B to treatment A decreases as $\mu_B$ increases. Yet this impact is counteracted by increasing the  certainty around $\omega_t$. Thus increasing time variability between times $t_{i-1}$ and $t_{i}$, indicates a larger necessary patient budget required to detect switching from treatment B to treatment A.

Figure 2 Here titled compweightA

\section*{Stopping Rule}
In an effort to keep this model fully Bayesian, a power analysis was conducted using a Bayes Factor, and the 95\% credible intervals along with the medians were calculated.
Determination of an appropriate Bayes Factor value has been described in \cite{kass1995bayes}, who indicate a Bayes Factor greater than 100 indicates Decisive evidence against the null hypothesis of no difference.

However, \cite{gonen2005bayesian} use the opposite notation for the Bayes Factor, whereby the null hypothesis is in the numerator yielding
\begin{equation}  \label{eq:8}
p(H_{0} \mid \textbf(D)) = \frac{P(\textbf{D} \mid H_{0}) P(H_{0}))} {P(\textbf{D} \mid H_{0})p(H_{0})+P(\textbf{D} \mid H_{1})p(H_{1})}  %\label{eq:8}
\end{equation}
whereby they have the null hypothesis in the numerator and this 
leads to the Bayes Factor
\begin{equation}  \label{eq:9}
BF_{01}= \frac{P(\textbf{D} \mid H_{0})}{P(\textbf{D} \mid H_{1})}
\end{equation}

which leads to their suggestion that a Bayes Factor less than $\frac{1}{100}$ provides decisive evidence against the null hypothesis and in favor of the alternative hypothesis.
The Bayes Factor was calculated using the Bayesian Two Sample T-Test discussed in \cite{gonen2005bayesian}. They define the Bayes Two Sample T Test as
\begin{equation}  \label{eq:10}
BF_{01} = \frac{T_{\nu}(t \mid 0,1)}{T_{\nu}(t\mid n_{\delta}^{\frac{1}{2}}\lambda,1+n_{\delta}\sigma_{\delta}^2)}
\end{equation}
The notation of \cite{gonen2005bayesian} was chosen as the more appropriate notation, and a stopping criterion was chosen to be a Bayes Factor of $\frac{1}{100}$, to provide \lq\lq{decisive evidence\rq\rq} and support towards the effective treatment. Any significant Bayes factor indicated a 100 times more likely chance the allocation had switched. Likewise, any indecisive Bayes Factor indicated the switch to the better treatment had not occurred. The bold numbers represent the Bayes Factor calculated at the budget size $\mathscr{N} = 100$ The values in parenthesis in Table~\ref{tbl:budget123} and Table~\ref{tbl:budget45} represent the median and 95\% credible interval values required to switch treatments.

%\begin{center}
\begin{table}[ht]
	\small\sf\centering
	\caption{Non Covariate Budget Allocation N using $\mu_{B} = 1, 2, 3$ ($Q_{0.025}$, $Q_{0.5}$, $Q_{0.975}$) $\mathbf{P(\mathscr{N} \ge 100)}$. Italicized values indicate Noteworthy Bayes Factor}
	\begin{center}
		\begin{adjustbox}{max width=\columnwidth}
		%	\resizebox{\columnwidth}
			\begin{tabular}{cc|rrr}\\
			\toprule
				$$&$$&$$&$\mu_{B}$&$$\\
			\midrule
				$C_t$&$\omega_t$&1&2&3\\
				%\hline
				\midrule
				\multirow{3}{*}{0.1}&0.1&(30, 52, 91), \textit{\textbf{0.007}}&(28, 39, 59), \textit{\textbf{0.000}}&(28, 37, 56), \textit{\textbf{0.001}}\\
				&0.01&(60, 89, 100), \textbf{0.319}&(64, 80, 100), \textbf{0.061}&(58, 72, 97.025), \textbf{0.018}\\
				&0.001&(100, 100, 100),\textbf{1.000}&(100, 100, 100),\textbf{0.985}&(75.975, 89.5, 100),\textbf{0.158}\\
				\hline
				\multirow{3}{*}{0.001}&0.1&(30, 51, 88), \textit{\textbf{0.007}}&(26, 45, 83), \textit{\textbf{0.005}}&(27, 36, 57), \textit{\textbf{0.001}}\\
				&0.01&(61, 89, 100), \textbf{0.303}&(64.975, 80, 100), \textbf{0.670}&(58, 72, 94), \textbf{0.011}\\
				&0.001&(100, 100, 100),\textbf{1.000}&(100, 100, 100),\textbf{1.000}&(98, 100, 100),\textbf{0.949}\\
				\hline
				\multirow{3}{*}{0.000001}&0.1&(29, 52, 88), \textit{\textbf{0.009}}&(28, 40, 60), \textit{\textbf{0.000}}&(28, 37, 54), \textit{\textbf{0.002}}\\
				&0.01&(61, 90, 100), \textbf{0.327}&(65, 80, 100), \textbf{0.071}&(59, 72, 95.025), \textbf{0.014}\\
				&0.001&(100, 100, 100),\textbf{1.000}&(100, 100, 100),\textbf{1.000}&(99, 100, 100),\textbf{0.973}\\
			\bottomrule
			\end{tabular}
		\end{adjustbox}
	\end{center}
	\label{tbl:budget123}
\end{table}
%\end{center}

\begin{table}[ht]
	\small\sf\centering
	\caption{Budget Allocation $\mathscr{N}$ using $\mu_{B} = 4, 5$ ($Q_{0.025}$, $Q_{0.5}$, $Q_{0.975}$) $\mathbf{P(N \ge 100)}$. Italicized values indicate Noteworthy Bayes Factor}
	\begin{center}
		\begin{adjustbox}{max width=\columnwidth}
			\begin{tabular}{cc| r r}
			\toprule 
				$$&$$&$\mu_{B}$$$\\ 
			\midrule
				$C_t$&$\omega_t$&4&5\\		
			\midrule
				\multirow{3}{*}{0.1}&0.1&(29, 39, 87.050), \textbf{0.013}&(31, 43, 100), \textbf{0.130}\\
				&0.01&(49.975, 61, 87), \textit{\textbf{0.004}}&(44, 56.5, 82.025), \textit{\textbf{0.003}}\\
				&0.001&(64, 75, 95), \textbf{0.012}&(57, 69, 88.025), \textit{\textbf{0.007}}\\
				\hline
				\multirow{3}{*}{0.1}&0.001&(29, 38, 76), \textbf{0.011}&(31, 42, 100), \textbf{0.111}\\
				&0.01&(50, 61, 82), \textit{\textbf{0.001}}&(45, 56, 77), \textit{\textbf{0.000}}\\
				&0.001&(87, 95, 100), \textbf{0.228}&(80, 89, 100), \textbf{0.049}\\
				\hline
				\multirow{3}{*}{0.000001}&0.1&(29, 38, 69.025), \textit{\textbf{0.006}}&(32, 42, 100), \textbf{0.104}\\
				&0.01&(50, 61, 79.025), \textit{\textbf{0.001}}&(45, 55, 74), \textit{\textbf{0.001}}\\
				&0.001&(86.975, 96, 100), \textbf{0.281}&(80, 89, 100), \textbf{0.060}\\
			\bottomrule	
			\end{tabular}
		\end{adjustbox}
	\end{center}
	\label{tbl:budget45}
\end{table}

Using $\mu_B = 1$ and   $\omega_t = 0.1$ and  $c_{t_B} = 0.000001$ it can be seen the 
median switch occurs at 52 (95\% credible interval 29, 88) with a decisive Bayes Factor value 0.009, indicating this was 100 times more likely to have switched to the favorable treatment. However, when $\omega_t = 0.01$ the median switch occurs at 90 (95\% credible interval 61, 100) with a indecisive Bayes Factor 0.327, indicating at $\mathscr{N} = 100$ the switch to the favorable treatment had not yet occurred. Finally, when $\omega_t = 0.001$, all quantiles were 100, with an indecisive Bayes Factor = 1.000 thereby indicating the more effective treatment had not yet been detected at $\mathscr{N} = 100$ and no switching had occurred.

Using $\mu_B = 3$ and   $\omega_t = 0.1$ and  $c_{t_B} = 0.000001$ median switch occurs at 37 (95\% credible interval 28, 54) with a decisive Bayes Factor 0.002, indicating this was 100 times more likely to have switched to the favorable treatment. However, when $\omega_t = 0.01$ the median switch occurs at 72 (95\% credible interval 59, 95.025) with a indecisive Bayes Factor 0.014, indicating at $\mathscr{N} = 100$ the switch to the favorable treatment had not yet occurred. Finally, when $\omega_t = 0.001$, the median switch occurs at 100 (95\% credible interval 99, 100) with a indecisive Bayes Factor 0.973, also indicating at $\mathscr{N} = 100$ the switch to the favorable treatment had not yet occurred. 

Lastly, using $\mu_B = 5$ and   $\omega_t = 0.1$ and  $c_{t_B} = 0.000001$ the median switch occurs at 42 (95\% credible interval 32, 100) with a indecisive Bayes Factor 0.104, indicating at $\mathscr{N} = 100$ the switch to the favorable treatment had not yet occurred. However, when $\omega_t = 0.01$ the median switch occurs at 55 (95\% credible interval 44, 74) with a decisive Bayes factor value 0.001 indicating this was 100 times more likely to have switched to the favorable treatment. Lastly, when $\omega_t = 0.001$ median switching value was 89 (95\% credible interval 80, 100) with an indecisive Bayes Factor value of 0.060, suggesting the switch to the favorable treatment had not occurred at $\mathscr{N} = 100$.

A careful examination of the remaining combinations indicates that for $\mu_{B}=$ 1, 2, and 3 the only decisive Bayes Factors $\omega_t = 0.1$, although the Bayes Factor does appear to diminish in these cases when $\omega_t = 0.01$, yet it remains indecisive. Likewise, at $\omega_t = 0.001$, the Bayes Factors are highly indecisive. However, when analyzing $\mu_{B} = 4$ the Bayes Factors for $\omega_t =$ 0.1 and 0.001 are decisive, while that for  $\omega_t = 0.001$ is indecisive.  Interestingly, $\mu_{B} = 4$ the scenario for $\omega_t = 0.001$ is the only decisive Bayes Factor. The behavior of these suggests if one wishes to investigate the impact of a smaller mean and seek definitive results, it is best to have lower certainty about the between time behavior and use $\omega_t = 0.1$, however, for the larger means a bit more certainty about between time variance $\omega_t = 0.01$ should be used to detect a decisive difference.

\section*{Conclusion}
Modern computational power has aided researchers by decreasing the amount of time necessary to run large simulations or large computationally difficult problems which may arise from when using Bayesian methods. Studies such as Bayesian adaptive designs in clinical trial 
benefit from this increased computational power through a decreased completion time, yet some Bayesian adaptive designs remain time consuming. The current application of the DLM to random allocation models illustrates its benefit through both greatly reduced allocation time and in decreased allocation size necessary to determine the most appropriate treatment.  
Likewise the corresponding sensitivity analysis illustrates the differing model behaviors and allocation proportions which one may expect to see when using the DLM to allocate patients to treatments. Finally the power analysis conducted provides users the ability to determine the proportion of available patient budget they may wish to use to determine appropriate stopping criterion. This should greatly reduce the number of ineffective treatment allocations and begin allowing the most effective treatment to be applied in a more timely manner through a smaller patient budget size.
However, the current application focuses only on random allocation models with no covariates therefore, the impact of a covariate such as gender or smoker was not included in this article and is something which will be addressed in a future article. Likewise, the possibility of a multi arm study is something which could be addressed in future work to determine if a particular treatment allocation can be removed from the study entirely. Additional future works may also include examining the Bayes factor stopping criterion from a survival analysis standpoint. \\

\bibliographystyle{SageV}
%bibliography{mybib7222020}

\end{document}